\documentclass[aps,preprint]{revtex4}
\usepackage{amsfonts}
\usepackage{amssymb}
\usepackage{amsmath}
\usepackage{graphicx}
\usepackage{hyperref}
\usepackage{float}
\usepackage{placeins}
\usepackage[font={footnotesize,it}]{caption}
\usepackage{latexsym}
\usepackage{xcolor}
\usepackage{hyperref}

\usepackage{bm}
\begin{document}

\title{Thermal Quantum Fluctuations in Einstein–Nonlinear Maxwell-Yukawa Black Hole}

\author{M. Mangut}
\email{mert.mangut@emu.edu.tr}
\affiliation{Department of Physics, Faculty of Arts and Sciences, Eastern Mediterranean
University, Famagusta, North Cyprus via Mersin 10, Turkey}

\begin{abstract}
This study investigates an Einstein–nonlinear electrodynamic black hole generated by a Yukawa-screened electromagnetic potential and examines its thermal properties through the calculation of the fundamental thermodynamic potentials and variables. The effect of thermal fluctuations on the calculated standard thermodynamic quantities is analyzed through two different correction terms added to the Bekenstein–Hawking entropy. In this work, the thermodynamic behavior in cases where the black hole has a relatively large event horizon is investigated using a logarithmic correction term representing corrections arising from quantum fields in constant backgrounds. In addition, thermal fluctuations occurring in cases where the event horizon is small are analyzed through an exponential correction term added to the entropy, reflecting non-perturbative quantum effects. Finally, the thermodynamic functions for all obtained cases are examined graphically, and the stability properties of the black hole are discussed in detail.
\end{abstract}

\maketitle

\section{Introduction}

The powerful integration of general relativity with the linear electromagnetic field represents one of the successful successes of classical field theory. In four-dimensional, static and spherically symmetric spacetimes, the Einstein–Maxwell equations admit the well-known Reissner–Nordström solution, describing a charged black hole uniquely determined by its mass and electric charge parameters \cite{reissner1916,nordstrom1918}. However, the standard solution provides a mathematically consistent model, the Coulomb interaction nature of the electric field introduces important limitations. In particular, the classical Coulomb field diverges at short distances, which does not accurately represent the behavior of charges in realistic physical contexts.

The limitations of linear Maxwell theory in regimes of high field strength have motivated the development of nonlinear electromagnetic models. In this context, Born and Infeld proposed the first nonlinear electromagnetic theory to regularize the self-energy divergence of the electromagnetic field \cite{borninfeld1934}. Subsequent developments in quantum electrodynamics, especially the effective Lagrangian formulated by Werner Heisenberg and Hans Euler, demonstrated that vacuum polarization naturally leads to nonlinear corrections to Maxwell’s equations \cite{heisenberg1936}. Overall, these developments indicate that the interaction between gravity and electromagnetic fields may be considerably more intricate than predicted by linear theory, potentially giving rise to new classes of black hole solutions and modified spacetime geometries. Also, coupling Einstein’s equations with nonlinear electromagnetic fields has produced remarkable solutions, such as those in \cite{ayonbeato1998,ayonbeato1999,bronnikov2001} which clearly show that nonlinearities in the electromagnetic sector can reshape both the singularity structure and horizon properties of black holes.

However, the natural interactions between charged particles are often short-ranged rather than truly long-ranged. Given the realistic nature of close interactions, it has been of scientific importance to develop a theoretical model that can explain them. Within this framework, Yukawa introduced a modification of the Coulomb potential to describe nuclear forces via the exchange of a massive mediator particle, leading to the well-known exponentially screened potential \cite{yukawa1935}. In fact, Yukawa’s insight has had profound implications not only in nuclear physics but also in broader theoretical contexts where short-range interactions are relevant, such as dense astrophysical and cosmological environments \cite{clifton2012}. Although the Yukawa gravitational potential has been extensively investigated in various contexts, including $f(R)$ modified gravity, cosmology, and galactic dynamics \cite{1,2,3,4,5,6,7,8}, a few studies, such as \cite{aa}, have also considered a Yukawa-like potential coupled to the electric field, deriving corresponding black hole solutions.

In  \cite{aa}, a spacetime geometry based on a black hole family obtained by generalizing the Coulomb potential to a Yukawa-type exponentially screened form is investigated, which consistently solved the coupled nonlinear electromagnetic and gravitational field equations  \cite{aa}. This solution can be interpreted as a short-range generalization of the classical Reissner–Nordström solution; however, it is not a simple potential modification. Rather, the resulting geometry emerges as a valid solution of the full field equations, reflecting the intricate interplay between short-range screening effects and nonlinear electromagnetic dynamics.

The physical relevance of the present analysis lies in determining how a finite-range nonlinear electromagnetic interaction modifies the thermodynamic response of charged black holes. Unlike the Reissner–Nordström solution, where the electromagnetic field extends over arbitrarily large distances, the Yukawa interaction becomes exponentially screened outside its characteristic length scale. Consequently, the influence of the electric charge is largely confined to the near-horizon region, while the spacetime gradually approaches Schwarzschild-like behavior at larger distances. This makes the Einstein–Maxwell–Yukawa solution an ideal theoretical laboratory for investigating how finite-range interactions affect black hole thermodynamics. In particular, the present work determines how finite-range electromagnetic screening modifies the Hawking temperature, thermodynamic potentials, heat capacity, and isothermal compressibility, thereby identifying the shifts in stability regions and phase-transition points relative to the classical Reissner–Nordström black hole. These results establish a direct connection between finite-range nonlinear electromagnetic interactions and observable thermodynamic signatures, demonstrating how Yukawa screening modifies the stability structure and phase behavior of charged black holes relative to the classical Reissner–Nordström solution.

From the geometric point of view,  the Yukawa parameter effectively suppresses the contribution of the electric field to the spacetime at large distances, significantly altering the horizon and causal structures as well as the thermodynamic properties of the black hole. In particular, the root structure of the metric function may qualitatively differ from that of the Reissner–Nordström solution, and extremal configurations no longer reduce to the classical $M = Q$ condition. This provides a natural framework to investigate the geometric and dynamical consequences of short-range field effects in strong gravity regimes, elucidating how deviations from idealized long-range interactions influence black hole physics. In addition, \cite{bb} investigated the de-Sitter generalization of the spacetime. Since Yukawa screening modifies the spacetime geometry and horizon structure, it is natural to ask whether these geometric changes are reflected in the thermodynamic response of the black hole. Addressing this question constitutes the primary objective of the present work.

Analyzing all aspects of the physics of proposed nonlinear models is crucial for demonstrating their theoretical advantage. In this context, analyzing the thermodynamics of the system, enabled by Hawking’s insight, allows us to draw important conclusions about its physical properties and to assess its consistency \cite{9,10,11}. In thermodynamic analyses, it is not sufficient to consider only the Hawking temperature and heat capacity in the context of stability. A more complete description requires the inclusion of internal energy, Helmholtz free energy, enthalpy, Gibbs free energy, and isothermal compressibility, which together provide deeper insight into the role of model parameters and physical consistency of the theory. Within this approach, the Hawking temperature and the general relativistic definition of entropy \cite{12,13} play the role of the fundamental starting points from which the thermodynamic potentials are derived. 

Additional contributions to entropy arising from different physical models and mathematical approaches, including corrections from various classical and quantum frameworks \cite{14,15,16}. These effects change not only the numerical values of thermodynamic quantities but also characterize the underlying microscopic structure of the system. Studying these corrections further clarifies how different theoretical frameworks influence stability criteria and thermodynamic potentials. This, in turn, provides a more complete picture of the model’s physical consistency and the limits within which the description remains valid. In this context, the generic corrections to the standard Hawking–Bekenstein entropy $(S_0)$ can be summarized as arising from small stable fluctuations around equilibrium \cite{17a} and non-perturbative quantum corrections \cite{17aa}. One of such contributions can be expressed as \cite{18}

\begin{equation}
S=S_0+\xi ln f_1(S_0)+\frac{\gamma}{S_0}+\eta e^{-S_0
}. \label{S}
\end{equation}

Here, $\xi,\gamma$ and $\eta$ are constants determined by the considered theoretical model, while $f_1(S_0)$ is function of $S_0$, which is defined by the mathematical structure of the model \cite{18}. In this work, the entropy corrections given in Eq.\eqref{S} can be examined separately from two different perspectives. The first approach is the case where the black hole has a sufficiently small horizon structure. In this limit, thermodynamic analyses can be performed by considering the entropy as $S \sim S_0 + \eta e^{-S_0}$ \cite{19}.

The second limit represents the case where the horizon radius of the black hole is relatively large. In this case, the general structure of entropy is reduced to the form $S \sim S_0 + \xi \ln f_1(S_0)$ \cite{20}. Also, thermodynamic analyses performed based on these two approaches for different spacetime geometries occupy a significant place in the literature \cite{a1,a2,a3,a,b,c,d,e,f,g}. Therefore, it is seen that under these two different limits, thermal fluctuations have significant effects on both thermodynamic potentials and stability analysis.

Although logarithmic and exponential entropy corrections have been considered for several black hole geometries, their physical consequences depend on the structure of the underlying gravitational and matter sectors. The purpose of the present work is therefore not to introduce these correction terms as new mathematical forms, but to determine how they interact with the finite-range nonlinear electromagnetic source of the Einstein–Maxwell–Yukawa black hole. In particular, the analysis allows us to distinguish corrections that primarily produce quantitative shifts from those that alter the qualitative behavior of the thermodynamic potentials and stability indicators. This comparison constitutes the main distinction between the present study and earlier investigations in which entropy corrections were applied to different black hole backgrounds.

The structure of the paper is organized as follows. In Section II, we briefly review the background geometry in the Einstein–Maxwell–Yukawa nonlinear electrodynamics (NLED) framework. Section III is devoted to the thermal structure of the corresponding black hole solution, where we analyze the standard thermodynamic behavior as well as logarithmic and exponential corrections to the entropy. In Section IV, we present and discuss our results.

\section{Review of the Background Geometry}
In \cite{aa}, a black hole solution sourced by Yukawa NLED was presented by Mazharimousavi and Halilsoy. 
Solving the field equations yields a static and spherically symmetric black hole geometry, with the general action given by

\begin{equation}
S=\frac{1}{16\pi G}\int d^4x\sqrt{-g}\left(R+\mathcal{L(F)} \right), \label{1}
\end{equation}

where $R$ is the Ricci scalar and $\mathcal{L(F)}$ denotes the nonlinear electromagnetic Lagrangian, depending on the Maxwell invariant $\mathcal{F}=F_{\mu\nu}F^{\mu\nu}$. To get one can vary the action  the nonlinear Maxwell equations,

\begin{equation}
\nabla_{\mu}\left(\frac{\partial \mathcal{L}}{\partial F} F^{\mu\nu}\right)=0, \label{2}
\end{equation}

while variation with respect to the metric tensor gives the Einstein field equations

\begin{equation}
G_{\mu\nu}=T_{\mu\nu}, \label{3}
\end{equation}

where the energy-momentum tensor of the nonlinear electromagnetic field reads

\begin{equation}
T_{\mu\nu}=\frac{1}{2}g_{\mu\nu}\mathcal{L}-2\frac{\partial \mathcal{L}}{\partial F} F_{\mu\lambda}F_{\nu}^{\ \lambda}. \label{4}
\end{equation}

Assuming a static and spherically symmetric spacetime, the line element can be written as

\begin{equation}
ds^{2}=-f(r)dt^{2}+\frac{dr^{2}}{f(r)}+r^{2}d\theta^{2}+r^{2}\sin^{2}\theta d\varphi^{2}, \label{5}
\end{equation}

where $f(r)$ represents the metric function of the background spacetime.

The electromagnetic four-potential corresponding to a purely electric field is taken as

\begin{equation}
A=\phi(r)\,dt, \label{6}
\end{equation}

where $\phi(r)$ denotes the time component of the electromagnetic four-potential and is assumed to have a Yukawa-modified Coulomb form:

\begin{equation}
\phi(r)=\frac{q}{r}e^{-\alpha r}, \label{7}
\end{equation}

where $q$ denotes the electric charge and $\alpha$ is a positive parameter controlling the range of the interaction. 
The corresponding electric field is obtained from

\begin{equation}
E(r)=-\frac{d\phi(r)}{dr}=\frac{q\left( 1+\alpha r\right)}{r^2 e^{\alpha r}}, \label{8}
\end{equation}

which allows the Maxwell invariant $\mathcal{F}$ to be explicitly expressed as a function of the radial coordinate $r$, providing a convenient form for further analytical computations.

In  \cite{aa}, the explicit form of the nonlinear electromagnetic Lagrangian for the Yukawa-modified field was also computed. The Lagrangian of fully characterizing the nonlinear electrodynamics sourcing the black hole solution is given by 

\begin{equation}
\mathcal{L}(\mathcal{F}) = \frac{C_0q}{r^4}\left\{\left(\alpha^3r^3\left(1+\alpha r\right)^2-1 \right)e^{-\alpha r}-\alpha^4r^4\mathcal{E}_{1}(\alpha r) \right\}+C_1. \label{9}
\end{equation}

Here, $C_0$ and $C_1$ are integration constants, while $\mathcal{E}_{1}(x)$ denotes the exponential integral, defined by

\begin{equation}
\mathcal{E}_{1}(x) = \int_1^\infty \frac{e^{-t}}{t} \, dt. \label{10}
\end{equation}

Solving the coupled Einstein–nonlinear electromagnetic field equations yields an exact black hole solution characterized by the mass parameter $M$, electric charge $q$, and Yukawa parameter $\alpha$.  The resulting metric function generalizes the standard Reissner–Nordström geometry and can be written as

\begin{equation}
f(r)=1-\frac{2M}{r}
-\frac{q^{2}\left(\alpha^{3}r^{3}-\alpha^{2}r^{2}+2\alpha r-6\right) e^{-\alpha r}}{6r^{2}}
+\frac{\alpha^{4}}{6}r^{2}q^{2}\mathcal{E}_{1}(\alpha r), \label{11}
\end{equation}

In order to recover the standard Reissner–Nordström solution in the limit $\alpha \to 0$, the integration constants are chosen as $C_1=0$ and $C_0=-q$, ensuring that the metric function reduces to the familiar RN form  \cite{aa}. Also note that, for $q\rightarrow0$ case, NLED metric \eqref{11} reduces to  well known Schwarzschild (Sch) metric.

 For nonvanishing $\alpha$, however, the exponential damping factor modifies the effective behavior of the electromagnetic field and leads to deviations from the standard charged black hole geometry, particularly in the near-horizon and short-distance regimes.
 
 Finally, in order to perform the thermodynamic analysis, the horizon structure of the black hole constitutes the fundamental basis of all thermal investigations. Therefore, it is necessary to examine the horizons of the solution given in Eq.\eqref{11} . However, due to the transcendental nature of the nonlinear solution under consideration, the determination and analytical investigation of its roots becomes technically difficult. In this context, the behavior of the horizons can be analyzed numerically by constructing a density plot that illustrates the variation of the roots with respect to the relevant parameters. To facilitate the numerical analysis and simplify the horizon structure, it is convenient to introduce dimensionless variables. By defining $\rho=\alpha r$, $Q=\alpha q$ and $m=\alpha M$ , the metric function can be rewritten in the following form
 
 \begin{equation}
f(\rho) = 1 - \frac{2 m}{\rho} 
- \frac{Q^2 \left(\rho^3 - \rho^2 + 2 \rho - 6\right) e^{-\rho}}{6 \, \rho^2} 
+\frac{Q^2 \, \rho^2}{6} \, \mathcal{E}_1(\rho). \label{12}
\end{equation}

\begin{figure}[h]
\centering
\includegraphics[width=10cm,trim=0 0 0 0,clip]{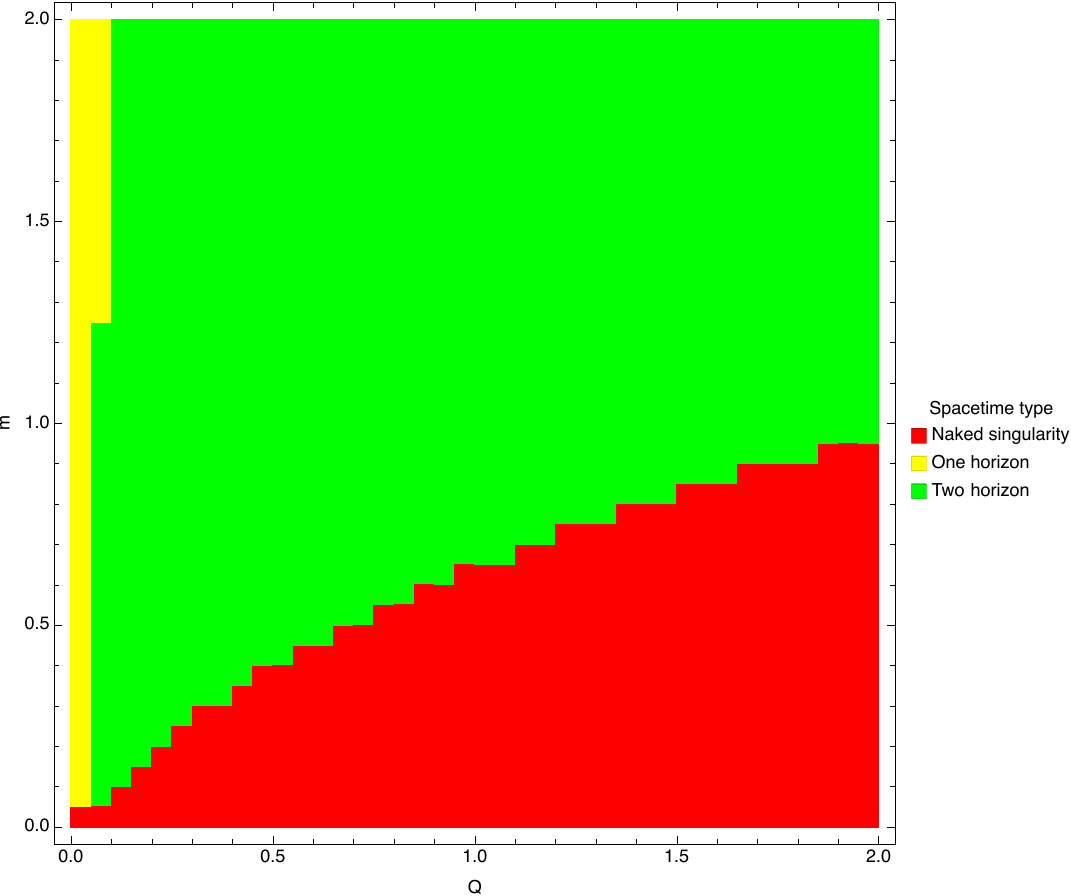}
\caption{This figure shows possible black hole solutions in the parameter space $(m,Q)$, where the density plot is generated by applying the condition $f(\rho)=0$ from Eq.\eqref{12}.}
\label{hi}
\end{figure}

To systematically investigate the horizon structure of the black hole, we employ a numerical approach and construct a density plot of the metric function $f(\rho)$ over the relevant parameter space. The resulting plot, shown in Fig.\ref{hi}, clearly illustrates how the locations and properties of the horizons depend on $m$ and $Q$.

The thermal analyses of the NLED metric, which will be examined in the next section, will be carried out  based on the metric form defined in Eq.\eqref{12}

\subsection{Effective Potential for Charged Timelike Test Particles}
We consider the motion of an electrically charged test particle with unit
mass and electric charge $q_p$ in the electrically charged Einstein--Nonlinear
Maxwell--Yukawa spacetime. The dynamics is described by
\begin{equation}
\mathcal{L}
=
\frac{1}{2}
g_{\mu\nu}
\dot{x}^{\mu}
\dot{x}^{\nu}
+
q_p A_{\mu}\dot{x}^{\mu},
\end{equation}
where an overdot denotes differentiation with respect to the proper time
$\tau$. The electromagnetic four-potential is
\begin{equation}
A_{\mu}
=
(\phi(r),0,0,0),
\end{equation}
with $\phi(r)$ given by Eq.\eqref{7}.
The stationarity and spherical symmetry of the spacetime imply the conserved
energy and angular momentum,
\begin{equation}
E
=
f(r)\dot{t}
-
q_pA_0
=
\mathrm{const},
\qquad
l
=
r^{2}\sin^{2}\theta\,\dot{\varphi}
=
\mathrm{const}.
\end{equation}
Owing to the spherical symmetry of the spacetime, the motion can be
restricted, without loss of generality, to the equatorial plane,

\begin{equation}
\theta=\frac{\pi}{2},
\qquad
\dot{\theta}=0.
\end{equation}
The conserved quantities then yield
\begin{equation}
\dot{t}
=
\frac{E+q_pA_0}
{f(r)},
\end{equation}
and
\begin{equation}
\dot{\varphi}
=
\frac{l}{r^{2}}.
\end{equation}
The timelike normalization condition,
\begin{equation}
g_{\mu\nu}\dot{x}^{\mu}\dot{x}^{\nu}=-1,
\end{equation}
together with the conserved quantities yields the radial equation
\begin{equation}
\left(E+q_pA_0\right)^{2}
=
\dot{r}^{\,2}
+
f(r)
\left(
1+\frac{l^{2}}{r^{2}}
\right).
\end{equation}
The effective potential governing the radial motion is defined by

\begin{equation}
2V_{\mathrm{eff}}
=
E^{2}
-
1
-
\dot{r}^{\,2},
\end{equation}
which yields
\begin{equation}
V_{\mathrm{eff}}
=
\frac{1}{2}(E^{2}-1)
-
\frac{1}{2}
\left[
\left(E+q_pA_0\right)^{2}
-
f(r)
\left(
1+\frac{l^{2}}{r^{2}}
\right)
\right].
\end{equation}
The effective potential provides a direct dynamical characterization of the
interaction between the charged test particle and the Yukawa-screened
electromagnetic background. In particular, its radial profile allows us to
examine how the nonlinear finite-range electromagnetic interaction modifies
the accessible regions and orbital behavior of charged particles.

\section{ Thermal Structure of Einstein-Maxwell-Yukawa  NLED Solution }

In this section, we analyze the standard thermodynamic properties of the Einstein-Maxwell-Yukawa (EMY) black hole solution given in Eq.\eqref{12} in the context of nonlinear electrodynamics. Firstly, the Hawking temperature, entropy, internal energy, thermodynamic volume, as well as the Helmholtz and Gibbs free energies are computed. In addition, the leading-order corrections to the thermodynamic potentials are determined. To analyze the thermodynamic stability, the heat capacity and the isothermal compressibility are also evaluated, as they play a crucial role in determining the stable and unstable regions of the black hole phase space under different correction mechanisms. \\

\subsection{Hawking Temperature and Classical Black Hole Thermodynamics}

Before including any quantum or higher-order corrections, we first compute the standard thermodynamic quantities of the black hole. The Hawking temperature is obtained from the horizon properties as follows

\begin{equation}
\begin{aligned}
T_H &= \left.\frac{1}{4\pi}\frac{\partial f(r)}{\partial r} \right|_{r=r_h}= \left.\frac{\alpha}{4\pi}\frac{\partial f(\rho)}{\partial \rho} \right|_{\rho=\rho_h}=\mathcal{T} \\
 &= \frac{\rho_h\left(\rho_h^3 Q^2 \mathcal{E}_1(\rho_h)+6m\right)
 -Q^2\left(\rho_h^3-\rho_h^2+2\rho_h+6\right)e^{-\rho_h}}
 {12\pi\rho_h^3},
\end{aligned}
\label{13}
\end{equation}

where $\mathcal{T}=\frac{T_H}{\alpha}$. If we apply the condition $f(\rho=\rho_h)=0$  in Eq.\eqref{12}, the mass function is expressed as

\begin{equation}
M(\rho_h,Q)=\frac{\rho_h}{2}-\frac{Q^2\left(\rho_h^3-\rho_h^2+2 \rho_h-6\right)  e^{-\rho_h }}{12 \rho_h}
+\frac{1}{12} \rho_h^3 Q^2 \mathcal{E}_1(\rho_h). \label{14}
\end{equation}

Substituting Eq.\eqref{14} into Eq.\eqref{13}, the Hawking temperature becames 

\begin{equation}
\mathcal{T}=\frac{ Q^2 \rho _h^4 \mathcal{E}_1\left(\rho _h\right)+2 \rho _h^2-Q^2 \left(\rho _h^3-\rho _h^2+2 \rho _h+2\right)e^{-\rho _h}}{8 \pi  \rho _h^3}. \label{15}
\end{equation}

\begin{figure}[H]
\centering
\includegraphics[width=10cm]{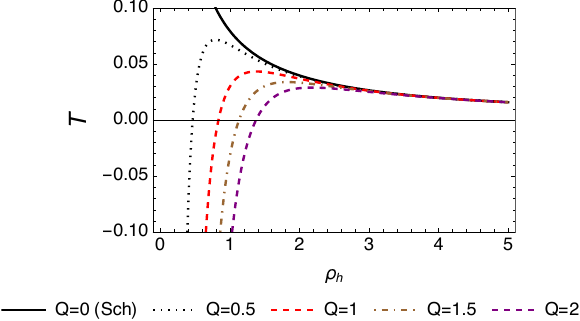}  
\caption{The Hawking temperature versus the dimensionless event horizon.}\label{f1}
\end{figure}

In Fig.\ref{f1}, we observe the behavior of the Hawking temperature for different values of the charge parameter. The effect of the charge is significant in the small dimensionless radius regime. However, in the large dimensionless radius regime, all cases converge and overlap with the Schwarzschild solution. Furthermore, the charge provides a notable contribution to the peak value of the temperature in the small radius regime.

For $G=1$, the classical Hawking–Bekenstein entropy $(S_0)$ of four-dimensional spherically symmetric black holes is defined by

\begin{equation}
S_0=\frac{1}{4}\int_0^{2\pi}\int_0^{\pi}\left.\sqrt{g_{\varphi\varphi}g_{\theta\theta}}\right |_{r=r_h}=\pi r^{2}_{h}. \label{16}
\end{equation}

In terms of the dimensionless radial coordinate $\rho=\alpha r$, the entropy is given by

\begin{equation}
\mathcal{S}_0=\alpha^2S_0=\pi\rho_h^2. \label{17}
\end{equation}

At this stage, we analyze the standard thermal properties of the nonlinear solution using Eqs.\eqref{15} and \eqref{16}. In this context, the  Helmhotz  free  energy , $\mathcal{F}$,  can be written as

\begin{equation}
\mathcal{F}=-\int \mathcal{S}_0d\mathcal{T}=-\int \frac{1}{8\rho_h^2} \left[ Q^2 \rho _h^4 \mathcal{E}_1\left(\rho _h\right)-2\rho_h^2+Q^2 \left(6+6\rho_h+\rho_h^2-\rho_h^3\right)e^{-\rho _h} \right]  d\rho_{h}, \label{18}
\end{equation}

and  the integration of Eq.\eqref{18}  gives

\begin{equation}
\mathcal{F}=\frac{\rho _h}{24}\left(Q^2 \rho _h^2 \left(E_{-2}\left(\rho _h\right)-E_1\left(\rho _h\right)\right)+6\right)-\frac{Q^2 e^{-\rho _h} \left(\rho _h^2-6\right)}{8 \rho _h} \label{19}.
\end{equation}

\begin{figure}[H]
\centering
\includegraphics[width=10cm]{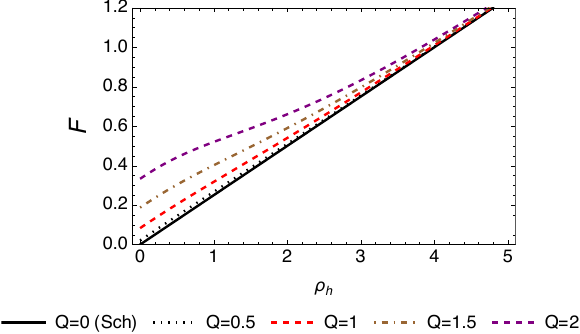}  
\caption{The graph shows the behavior of the Helmhotz  free  energy for different values of the dimensionless radius parameter.}\label{f2}
\end{figure}

Fig.\ref{f2}  shows that the Helmholtz free energy increases with the charge parameter, indicating a positive contribution of the charge to the system’s total energy and thermodynamic potential in the small dimensionless radius regime. However, this effect disappears as the dimensionless radius increases.

The generic internal  energy   of  the  system  is formulated by 

\begin{equation}
\mathcal{U}=\int \mathcal{T} d\mathcal{S}_0. \label{20}
\end{equation}

Substituting the corresponding expressions of temperature \eqref{15} and entropy \eqref{17} into equation \eqref{20}, we obtain

\begin{equation}
\mathcal{U}=\frac{1}{12} Q^2 \rho _h^3 \left(\mathcal{E}_1\left(\rho _h\right)-\mathcal{E}_{-2}\left(\rho _h\right)\right)+\frac{ \rho _h}{2}+\frac{Q^2 e^{-\rho _h} \left(\rho _h^2+2\right)}{4\rho _h} \label{21}.
\end{equation}

\begin{figure}[H]
\centering
\includegraphics[width=10cm]{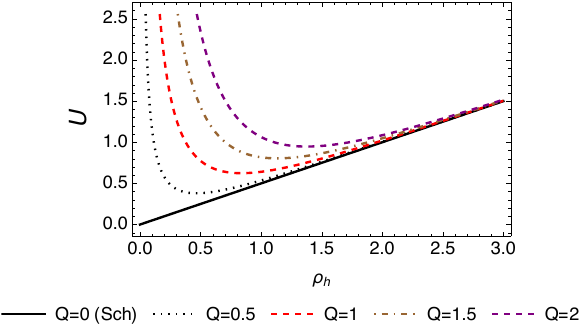}  
\caption{Behavior of the internal energy with respect to the dimensionless horizon radius for different charge values.} \label{f3}
\end{figure}

In Fig.\ref{f3} the small dimensionless horizon regime, the internal energy increases with increasing charge, following an exponential behavior. As the horizon radius increases, this behavior  approaches a linear regime and overlaps with it. Thermally, this indicates that charge contributions to the energy dominate at small horizons, whereas at large horizons the system approaches a stable classical regime.

The general expression for the pressure is given by

\begin{equation}
P=-\frac{dF}{dV}. \label{22a}
\end{equation}

\begin{figure}[H]
\centering
\includegraphics[width=10cm]{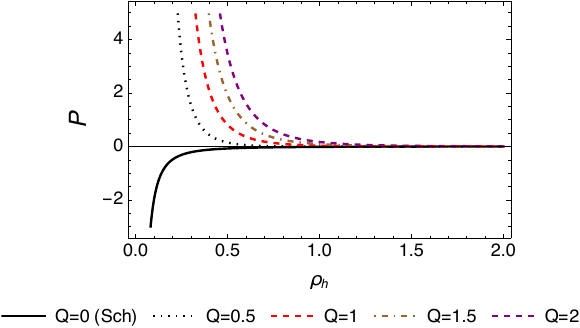}  
\caption{Pressure versus dimensionless black hole horizon for various $Q$ values.}\label{f4}
\end{figure}

Here, $V$ represents the standard thermodynamic volume of the black hole, defined as $V=\frac{4}{3}\pi r_h^3$. Plugging Eq.\eqref{19} into Eq.\eqref{22a}, the pressure of the black hole becomes

\begin{equation}
\mathcal{P}=-\frac{d\mathcal{F}}{d\mathcal{V}}=\frac{Q^2 \left(-\rho _h^3+\rho _h^2+6\rho _h+6\right)e^{-\rho _h} +\rho _h^2 \left(Q^2 \rho _h^2 \mathcal{E}_1\left(\rho _h\right)-2\right)}{32 \pi  \rho _h^4} \label{22}
\end{equation}

in which $\mathcal{V}=\alpha^3V$. Fig.\ref{f4} shows that increasing the charge shifts the pressure curve upward, indicating that higher charge increases the pressure for a given dimensionless horizon radius. For the uncharged case $(Q=0)$, corresponding to the Schwarzschild limit, the pressure is lower compared to charged cases and the curve shows a monotonic behavior. 

\begin{figure}[H]
\centering
\includegraphics[width=10cm]{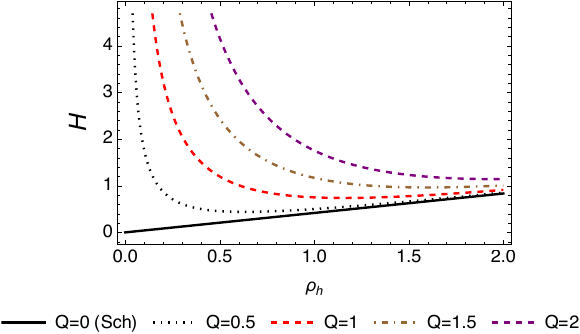}  
\caption{Enthalpy behaviors versus the dimensionless black hole horizon for different $Q$ values.}\label{f5}
\end{figure}

The enthalpy $\mathcal{H}$  is commonly defined as $\mathcal{H}=\mathcal{U}+\mathcal{P}\mathcal{V}$; in the scenario considered here, it can be expressed as

\begin{equation}
\mathcal{H}=\frac{3 Q^2 \rho _h^4 \left(\text{Shi}\left(\rho _h\right)-\text{Chi}\left(\rho _h\right)\right)+10 \rho _h^2+Q^2 e^{-\rho _h} \left(\rho _h \left(2-3 \left(\rho _h-1\right) \rho _h\right)+18\right)}{24 \rho _h},     \label{23}
\end{equation}

where the hyperbolic sine and cosine integrals are defined by
\begin{equation}
\operatorname{Shi}(x)=\int_{0}^{x}\frac{\sinh t}{t}\,dt,
\qquad
\operatorname{Chi}(x)=\gamma+\ln x+\int_{0}^{x}\frac{\cosh t-1}{t}\,dt,
\end{equation}
in which \(\gamma\) is the Euler–Mascheroni constant. In Fig.\ref{f5}, it is observed that the enthalpy shows the same characteristic behavior as the internal energy, as both are governed by the same underlying thermodynamic structure. In this context, enthalpy accounts for both the internal energy and the pressure–volume work of the system.

The Gibbs free energy, formulated in terms of the relevant thermodynamic variables, is given by

\begin{equation}
\mathcal{G}=\mathcal{F}+\mathcal{P}\mathcal{V}. \label{24g}
\end{equation}

When we substitute the relevant thermodynamic variables into Eq.\eqref{24g}, the Gibbs free energy of the black hole reduces to

\begin{equation}
\mathcal{G}= \frac{\rho _h}{6}+\frac{Q^2 e^{-\rho _h} \left(\rho _h+3\right)}{3 \rho _h}    \label{24}
\end{equation}

\begin{figure}[H]
\centering
\includegraphics[width=10cm]{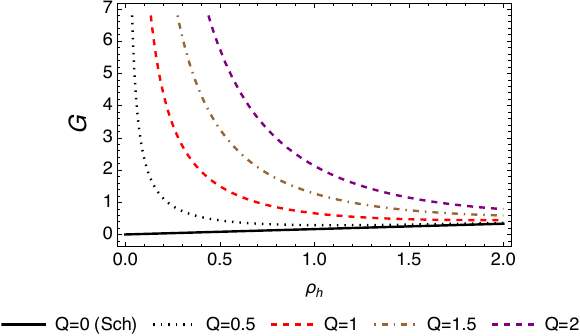}  
\caption{The graph of the Gibbs free energy with respect to the dimensionless black hole horizon with various  $Q$ choices.} \label{f6}
\end{figure}

As shown in Fig.\ref{f6}, the graphical structure is closely similar to that of the enthalpy. An increase in the charge parameter leads to similar monotonic behavior in the energy for both small and large dimensionless horizon values. 

The isothermal compressibility, which ensures that the system returns to equilibrium under spontaneous parameter changes in accordance with Le Chatelier’s principle, is defined as

\begin{equation}
\kappa=-\frac{1}{V}\frac{\partial V}{\partial p} \bigg\vert_T. \label{25}
\end{equation}

By substituting Eq.\eqref{22} into Eq.\eqref{25} applying the chain rule, the isothermal compressibility can be expressed as

\begin{equation}
\mathcal{\kappa}=\frac{24 \pi  e^{\rho _h} \rho _h^4}{2 Q^2 \left(\rho _h^2 +3\rho _h+3\right)-e^{\rho _h} \rho _h^2} \label{26}
\end{equation}

\begin{figure}[H]
\centering
\includegraphics[width=10cm]{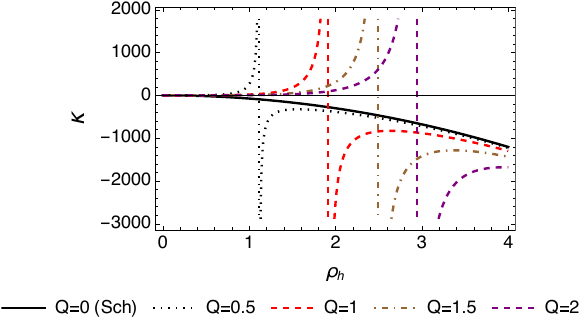}  
\caption{The structure of the $\kappa$  plots is ploted by Eq.\eqref{26}. }\label{f7}
\end{figure}

\begin{figure}[H]
\centering
\includegraphics[width=10cm]{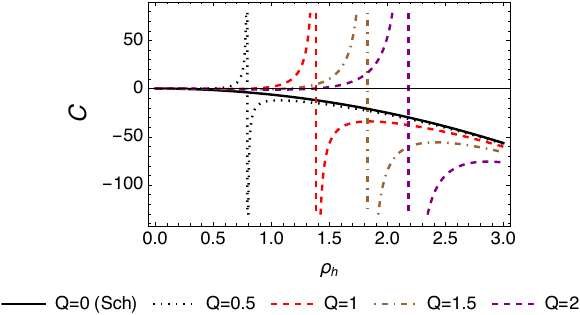}  
\caption{The plot of $\mathcal{C}$ are governed by Eq.\eqref{27}.} \label{f8}
\end{figure}

In Fig.\ref{f7}, the divergence of the isothermal compressibility at specific points shows the presence of critical behavior, where the system becomes infinitely sensitive to pressure fluctuations, indicating a phase transition. With increasing $Q$, these divergence points shift toward higher values of $\rho_h$. In the Sch case $(Q=0)$, the isothermal compressibility exhibits a monotonic decrease and remains negative over the entire domain, with no divergence, in analogy with the behavior of the heat capacity (See Fig.\ref{f8}).

The heat capacity is an important thermodynamic quantity that indicates how the system’s energy changes with temperature. In black hole thermodynamics, it provides insight into stability: a positive heat capacity corresponds to a stable configuration, while a negative value characterizes instability. For EMY NLED case, the heat capacity is given by

\begin{equation}
\mathcal{C}=\frac{d\mathcal{U}}{d\mathcal{T}}= \frac{2 \pi \rho _h^2 \left[2\rho _h^2+Q^2 \rho _h^4\mathcal{E}_1\left(\rho _h\right)-Q^2 \left(2+2 \rho _h-\rho _h^2+ \rho _h^3\right)e^{-\rho _h}\right]}{-2\rho _h^2+Q^2 \rho _h^4\mathcal{E}_1\left(\rho _h\right)-Q^2 \left(-6-6 \rho _h-\rho _h^2+ \rho _h^3\right)e^{-\rho _h}}. \label{27}
\end{equation}

The plots showed in Fig.\ref{f8} represent the behavior of $\mathcal{C}$ as a function of the dimensionless black hole horizon $\rho_h$, for different values of charge parameter $Q$. The important structures on the graph are summarized below.

    \begin{itemize}
        \item At  $Q=0$, $\mathcal{C}$ continuously decreases as $\rho_h$ increases, representing a standard trend.
        \item At $Q=0.5,1,1.5$ and $2$, the heat capacity $\mathcal{C}$ develops divergences at specific points. It is worth noting that, with increasing $Q$, the corresponding divergence points shift to higher values of $\rho_h$.
    \end{itemize}

In contrast to Fig.\ref{f8}, the thermodynamic properties of the charged black hole in the presence of the Yukawa field reveal critical behavior in the specific heat $\mathcal{C}$, which diverges at particular points, indicating second-order phase transitions. These results point to a unique thermodynamic structure for such black holes, in which standard phase-transition descriptions may not be entirely applicable. Nonetheless, the presence of critical points in $\mathcal{C}$ highlights the system’s complex thermodynamic character, motivating further investigation.

\subsection{Logarithmic Corrections to Black Hole Entropy}

Kaul and Majumdar \cite{log} derived an exact formula for the entropy of a four-dimensional nonrotating black hole within the framework of quantum geometry, where the event horizon is described by boundary states of an $\mathrm{SU}(2)_k$ Chern-Simons theory. In the large horizon area limit, they obtained the leading semiclassical Bekenstein-Hawking term proportional to the area, along with a subleading logarithmic correction arising from quantum fluctuations of the horizon geometry. Their main result is given by

\begin{equation}
S_{\mathrm{bh}} = S_{\mathrm{0}} - \frac{3}{2} \ln \left( \frac{S_{\mathrm{0}}}{\ln 2} \right) + \text{constant} + \mathcal{O}(S_{\mathrm{BH}}^{-1}), \label{28}
\end{equation}

Here, the coefficient $-\frac{3}{2}$ of the logarithmic term is universal, independent of the Barbero-Immirzi parameter and unaffected by the inclusion of higher spin punctures or variations in the Chern-Simons level $k$. This quantum correction arises from nonperturbative fluctuations of spacetime geometry itself, in contrast to corrections from quantum matter fields in fixed backgrounds. The factor $\ln 2$ contributes only an additive constant and can therefore be absorbed and omitted without affecting the leading behavior of the entropy. In addition, subleading corrections of order $\mathcal{O}(S_0^{-1})$ are suppressed in the large horizon area limit and are neglected. Consequently, the simplified form adopted in this work is given by

\begin{equation}
S_{LC} \approx S_{0} - \frac{3}{2} \ln \left( S_{0} \right). \label{29}
\end{equation}

Based on the logarithmically corrected entropy \eqref{29}, the thermodynamic analysis can be proceeded with. In this framework, the logarithmic corrected Helmhotz  free  energy ($\mathcal{F}_{LC}$) can be written as 

\begin{equation}
\mathcal{F}_{LC}=-\int \mathcal{S}_{LC}d\mathcal{T} . \label{30}
\end{equation}

Substituting the Hawking temperature and logarithmic corrected entropy $(S_{\mathrm{LC}})$ into Eq.\eqref{30}, can be obtained

\begin{align}
\mathcal{F}_{LC}=\frac{e^{-\rho_h}}{48 \pi \rho_h^3} \Bigg(
& 2 \pi Q^2 e^{\rho_h} \rho_h^6 
\left( \mathcal{E}_{-2}(\rho_h) - \mathcal{E}_1(\rho_h) \right) \notag \\
& + 3 e^{\rho_h} \rho_h^4 
\left( 3 Q^2 \mathcal{E}_1(\rho_h) (2 \log\rho_h - 1) + 4\pi \right) \notag \\
& - 9 Q^2 \rho_h^3 
\left( e^{\rho_h} \mathcal{E}_1(\rho_h) + e^{\rho_h} \mathcal{E}_2(\rho_h)\log\pi + 2\log\rho_h - 1 \right) \notag \\
& + 6 \rho_h^2 
\left( 6 e^{\rho_h} + 3 e^{\rho_h} \log(\pi \rho_h^2) - Q^2 \right) \notag \\
& + 12 Q^2 \rho_h + 12 Q^2
\Bigg). \label{31}
\end{align}

\begin{figure}[H]
\centering
\includegraphics[width=10cm]{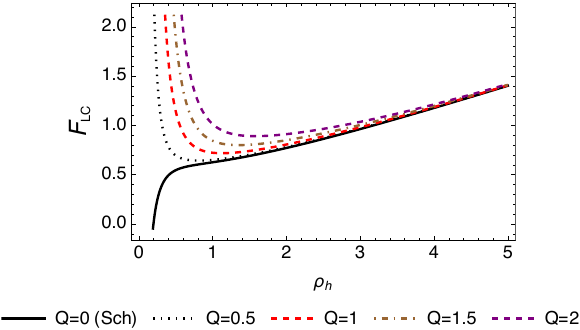}  
\caption{Variation of the logarithmically corrected Helmholtz free energy versus the dimensionless horizon radius $\rho_h$ with different values of the charge parameter $Q$.} \label{f9}
\end{figure}

Fig.\ref{f9} shows the variation of $\mathcal{F}_{LC}$ with the dimensionless event horizon radius $\rho_h$ for different charge parameters $Q$. The logarithmically corrected Helmholtz free energy increases monotonically, starting from low values in the small $\rho_h$ region for all $Q$ values. Using the Schwarzschild case ($Q=0$) as a reference, it is clearly seen that the curves shift upwards with increasing charge parameter. However, the effect of the logarithmic correction can be clearly observed in the region of small dimensionless horizon parameter because, unlike the case without correction, it exhibits a rapid decrease from a high value, followed by merging beyond the small horizon regime. Also, this behavior implies that the contribution of $Q$ is dominant at small dimensionless horizon radius but becomes subleading as the system size grows.

\begin{figure}[H]
\centering
\includegraphics[width=10cm]{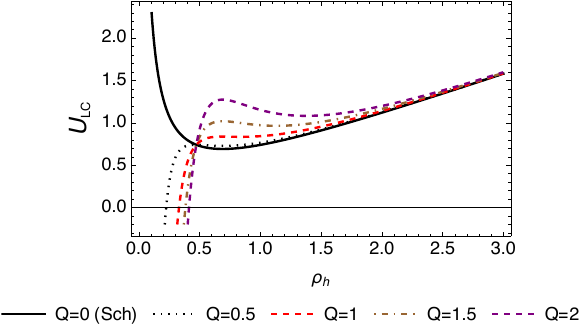}  
\caption{The graph of $\mathcal{U}_{LC}$ versus the dimensionless horizon radius $\rho_h$ with different charge cases.} \label{f10}
\end{figure}

In the light of new correction perspective, the corrected  internal  energy  expression 

\begin{equation}
\mathcal{U}_{LC}=\int \mathcal{T} d\mathcal{S}_{LC},  \label{32}
\end{equation}

can be written as

\begin{align}
\mathcal{U}_{LC}=\frac{1}{8\pi}\Bigg(
& -\frac{2}{3}\pi Q^2 \rho_h^3 \mathcal{E}_{-2}(\rho_h)
+ \frac{2}{3}\pi Q^2 \rho_h^3 \mathcal{E}_1(\rho_h)
+ 3Q^2 \mathcal{E}_2(\rho_h) \notag \\
& + 8Q^2 \mathrm{Ei}(-\rho_h)
+ 4\pi \rho_h
+ \frac{6}{\rho_h} \notag \\
& + \frac{Q^2 e^{-\rho_h}\left(2\pi \rho_h^4 + (5+4\pi)\rho_h^2 - 2\rho_h - 2\right)}{\rho_h^3}
\Bigg),  \label{33}
\end{align}

in which $\mathrm{Ei}$ represents  the exponential integral function. Fig.\ref{f10} represents the evaluation of logarithmically corrected internal energy $\mathcal{U}_{LC}$ with respect to the dimensionless horizon radius $\rho_h$. For all curves, the internal energy exhibits a regular and approximately linear increase with increasing $\rho_h$. This behavior is an expected physical consequence, showing that the total energy of the system increases as its size increases. As the charge parameter $Q$ increases, the curves shift upwards in the small $\rho_h$ regime. This clearly shows that the charge provides an additional energy contribution to the system. The Schwarzschild case corresponds to the lowest energy, while the highest $Q$ value gives the highest energy curve. In other words, $Q$ value gives the highest energy curve. However, the interesting difference here is that in the $\rho_h<0.5$ region, the charge contribution reverses its role and increases, thus reducing energy.

For logarithmic corrected thermodynamics analysis, the generic pressure formula \ref{22a} is  modified by

\begin{equation}
\mathcal{P}_{LC}=-\frac{d\mathcal{F}_{LC}}{d\mathcal{V}}, \label{34}
\end{equation}

and yields

\begin{equation}
\begin{aligned}
\mathcal{P}_{LC}=\frac{e^{-\rho_h}}{64 \pi^2 \rho_h^6}
\Bigg(
 \rho_h^3 \Big[
e^{\rho_h}\rho_h Q^2(-\mathcal{E}_1(\rho_h))
\big(-2\pi \rho_h^2 + 6\log\rho_h + 3 + 3\log\pi\big)
- 4\pi \rho_h^2 \\
\quad + 3e^{\rho_h}\rho_h^2 \log(\pi\rho_h^2)
+ Q^2(3\rho_h^2 + 2\rho_h + 12)
\Big]
+ 12Q^2(1+\rho_h)
\Bigg) .
\label{35}
\end{aligned}
\end{equation}

\begin{figure}[H]
\centering
\includegraphics[width=10cm]{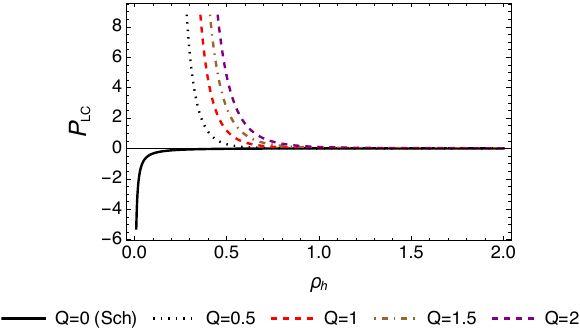}  
\caption{The plot of $\mathcal{P}_{LC}$ are governed by Eq.\eqref{35}.}\label{f11}
\end{figure}

As clearly seen in Fig.\ref{f11}, the logarithmic correction does not produce any significant change in the pressure trend; however, it exhibits the same behavior, starting from higher pressure values compared to the uncorrected case.

In this case  the logarithmic corrected  enthalpy ($\mathcal{H}_{LC}=\mathcal{U}_{LC}+\mathcal{P}_{LC}\mathcal{V}$)  can be found as

\begin{equation}
\begin{aligned}
\mathcal{H}_{LC}=\frac{e^{-\rho_h}}{48 \pi \rho_h^3}
\Bigg[
\rho_h^3 \Big(
3 Q^2 e^{\rho_h} \mathcal{E}_1(\rho_h)
\Big(2 \rho_h (2 \log \rho_h - 2 + \log \pi) - 3 \Big)
+ 8 \pi e^{\rho_h} \rho_h \\
\quad + Q^2 \Big(2 \pi \rho_h (\rho_h + 2)
- 18 \log \rho_h + 4 \pi + 10 - 9 \log \pi \Big)
\Big) \\
+ 12 e^{\rho_h} \big(2 \log(\pi \rho_h^2) + 3 \big)
- 4 Q^2 \rho_h
+ 24 Q^2 (\rho_h + 1)
\Bigg]\label{36}
\end{aligned}
\end{equation}

\begin{figure}[H]
\centering
\includegraphics[width=10cm]{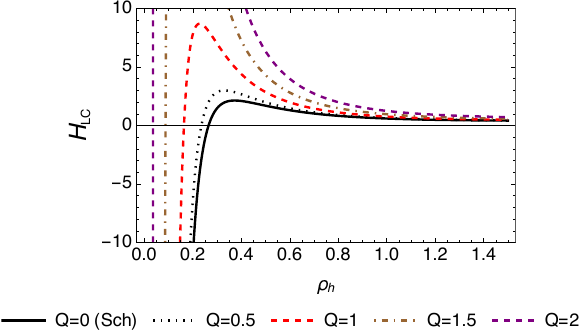}  
\caption{ The logarithmic corrected  enthalpy versus the dimensionless event horizon graphs according to Eq.\eqref{36} for different $Q$ values.}\label{f12}
\end{figure}

As illustrated in Fig.\ref{f12}, in the regime of large dimensionless horizon radius the system behaves similarly to the uncorrected enthalpy. However, in the small dimensionless horizon limit, the effect of logarithmic corrections becomes pronounced: the quantity decreases logarithmically from its maximum peak value as the dimensionless horizon radius increases, eventually approaching a linear behavior. Moreover, the maximum values of these peaks increase proportionally with the charge parameter.

Plugging Eq.\eqref{31} and Eq.\eqref{35} into Eq.\eqref{24g}, one finds

\begin{equation}
\begin{aligned}
\mathcal{G}_{LC}=\frac{e^{-\rho_h}}{48 \pi \rho_h}
\Bigg[
\rho_h^2 \Big(
3 Q^2 e^{\rho_h} \mathcal{E}_1(\rho_h)
\big(2 \pi \rho_h^2 - 2 \log \rho_h - 7 - \log \pi\big)
+ 20 \pi e^{\rho_h} \\
\quad - 4 \pi Q^2 (\rho_h - 1)
\Big)
+ Q^2 \big(48 e^{\rho_h} E_1(\rho_h) - 8 \pi + 19\big) \\
+ 6 e^{\rho_h} \big(\log(\pi \rho_h^2) + 6\big)
+ 8 (4 + 3\pi) Q^2
\Bigg]\label{37}.
\end{aligned}
\end{equation}

\begin{figure}[H]
\centering
\includegraphics[width=10cm]{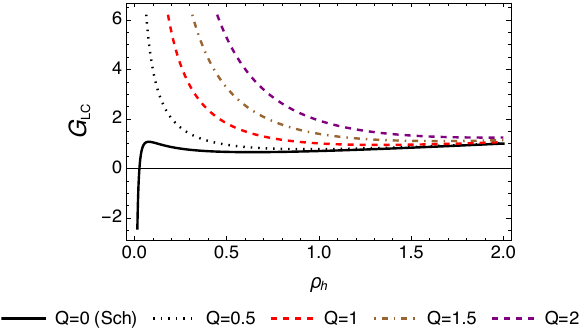}  
\caption{Behavior of the logarithmically corrected Gibbs free energy for different charge parameters.}\label{f13}
\end{figure}

As is clearly seen in Fig.\ref{f13}, there is no noticeable difference between the modified Gibbs free energy, which includes corrections, and the unmodified Gibbs free energy. However, in the absence of charge, a sudden turning point appears around $\rho_h\sim0.1$. This indicates that the logarithmic correction leads to a distinct deviation in the Sch case.

Upon the substitution of Eq.\eqref{35}  in the generic definition of isothermal compressibility \eqref{25}, one obtains

\begin{equation}
\begin{aligned}
\mathcal{\kappa}_{LC}
= -192 \pi^2 e^{\rho_h} \rho_h^6
\Bigg[
& 2 e^{\rho_h} \rho_h^2
\Big(
\rho_h^2 \big(
3 Q^2 \mathcal{E}_1(\rho_h) (2 \log \rho_h + \log \pi)
+ 4 \pi
\big) - 12 \log(\pi \rho_h^2) + 6
\Big)
\Bigg. \\
& \Bigg.
- Q^2 \Big(
\rho_h^4 (2 \pi \rho_h - 6 \log \rho_h - 2 - 3 \log \pi)
+ 5 \rho_h^3 + 20 \rho_h^2 + 72 \rho_h + 72
\Big)
\Bigg]^{-1} \label{38}
\end{aligned}
\end{equation}

\begin{figure}[H]
\centering
\includegraphics[width=10cm]{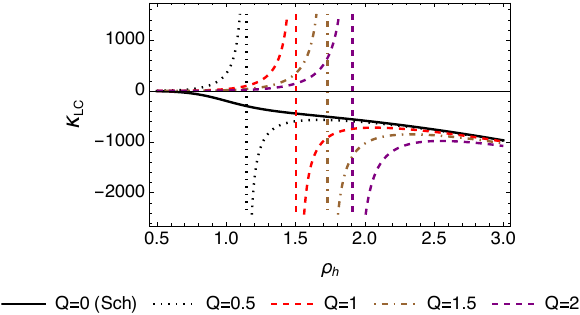}  
\caption{Isothermal compressibility versus dimensionless black hole horizon for various Q values.}\label{f14}
\end{figure}

It is observed that in Fig.\ref{f14} the inclusion of corrections does not introduce any fundamental change in the mathematical structure of the isothermal compressibility. However, for different charge values, the divergence points are confined to the interval between $1$ and $2$. This indicates that the logarithmic correction shifts the possible phase transition points slightly toward lower dimensionless horizon radius values.

To define the heat capacity with logarithmic corrections, one needs to use

\begin{equation}
\mathcal{C}_{LC}=\frac{d \mathcal{U}_{LC}}{d\mathcal{T}}\label{39a}.
\end{equation}

Hence we obtain the logarithmically corrected heat capacity as 

\begin{equation}
\begin{aligned}
&\mathcal{C}_{LC}=\frac{2}{3} \pi Q^2 \rho_h^3 \mathcal{E}_{-3}(\rho_h)
-2 \pi Q^2 \rho_h^2 \mathcal{E}_{-2}(\rho_h)
+2 \pi Q^2 \rho_h^2 \mathcal{E}_1(\rho_h)
-3 Q^2 \mathcal{E}_1(\rho_h)
-\frac{6}{\rho_h^2}
-\frac{2}{3} \pi Q^2 e^{-\rho_h} \rho_h^2 \\
&+\frac{Q^2 e^{-\rho_h}(8\pi \rho_h^3+2(5+4\pi)\rho_h-2)}{\rho_h^3}
-\frac{Q^2 e^{-\rho_h}(2\pi \rho_h^4+(5+4\pi)\rho_h^2-2\rho_h-2)}{\rho_h^3} \\
&-\frac{3Q^2 e^{-\rho_h}(2\pi \rho_h^4+(5+4\pi)\rho_h^2-2\rho_h-2)}{\rho_h^4}
+\frac{8Q^2 e^{-\rho_h}}{\rho_h}
+4\pi \\
&\quad \times \Bigg[
8\pi
\Bigg(
\frac{e^{-\rho_h}(Q^2 e^{\rho_h}\rho_h^4 \mathcal{E}_1(\rho_h)
+4Q^2 e^{\rho_h}\rho_h^3 \mathcal{E}_1(\rho_h)
+2e^{\rho_h}\rho_h^2
+4e^{\rho_h}\rho_h
-Q^2\rho_h^3
-Q^2(3\rho_h^2-2\rho_h+2))}{8\pi\rho_h^3} \\
&\qquad
-\frac{e^{-\rho_h}(Q^2 e^{\rho_h}\rho_h^4 \mathcal{E}_1(\rho_h)
+2e^{\rho_h}\rho_h^2
-Q^2(\rho_h^3-\rho_h^2+2\rho_h+2))}{8\pi\rho_h^3} \\
&\qquad
-\frac{3e^{-\rho_h}(Q^2 e^{\rho_h}\rho_h^4 \mathcal{E}_1(\rho_h)
+2e^{\rho_h}\rho_h^2
-Q^2(\rho_h^3-\rho_h^2+2\rho_h+2))}{8\pi\rho_h^4}
\Bigg)
\Bigg]^{-1} \label{39}
\end{aligned}
\end{equation}

\begin{figure}[H]
\centering
\includegraphics[width=10cm]{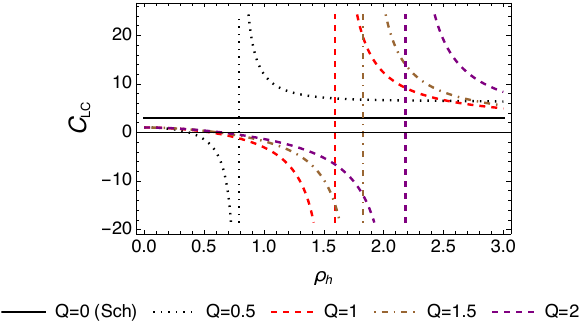}  
\caption{Logarithmically corrected heat capacity $\mathcal{C}_{LC}$ as a function of $\rho_h$ for various values of $Q$, illustrating the effect of logarithmic corrections on thermodynamic behavior.}\label{f15}
\end{figure}

Figure \ref{f15} illustrates the behavior of the logarithmically corrected heat capacity $\mathcal{C}_{LC}$ as a function of the dimensionless horizon parameter $\rho_h$, for different values of the charge parameter $Q$, including the Sch case $Q=0$. As the charge parameter $Q$ increases, the behavior of $\mathcal{C}_{LC}$ changes significantly. For nonzero $Q$, the curves exhibit divergences at certain critical values of $\rho_h$. These divergences correspond to second-order phase transition points, where the heat capacity changes sign. Physically, such points separate thermodynamically unstable regions (negative $\mathcal{C}_{LC}$) from stable ones (positive $\mathcal{C}_{LC}$). Moreover, the location of the divergence points shifts with increasing $Q$, indicating that the critical radius depends sensitively on the charge parameter. This reflects the nontrivial interplay between gravitational and electromagnetic contributions to the black hole thermodynamics. Also note that, When logarithmic corrections are included, it is observed that the phase transition points corresponding to the given $Q$ values shift further to the left compared to the case without corrections and for Sch limit $(Q=0)$ with logarithmic corrections, the heat capacity does not remain negative throughout the domain but instead attains a constant value.

\subsection{Exponential Corrections to Black Hole Entropy}

The exponential correction to black hole entropy can be motivated from a statistical description of horizon microstates. We consider a system of indistinguishable micro-particles, for which the total number of configurations is given by \cite{19}

\begin{equation}
\Omega = \frac{\left(\sum_i s_i\right)!}{\prod_i s_i!}. \label{40}
\end{equation}

Standard combinatorial arguments, together with constraints on the total particle number and energy, lead to the leading-order entropy contribution $S_0 = \lambda N$ \cite{19}, where $s_i$ denotes the occupation number of the $i$-th microstate, $N$ is the total number of micro-particles, and $\lambda$ is the Lagrange multiplier associated with the particle-number constraint, which plays the role of an effective thermodynamic parameter. To account for subleading effects in the microstate counting, often associated with non-perturbative contributions in quantum gravitational systems, we adopt the corrected form \cite{19}

\begin{equation}
S^{EC} = S_0 + e^{-S_0}. \label{41}
\end{equation}

If one inserts Eq.\eqref{41} into Eq.\eqref{30}, the exponentially corrected Helmhotz  free  energy can be explicitly written as

\begin{equation}
\begin{aligned}
\mathcal{F}_{EC}=&\frac{1}{24 \pi \rho_h^3}
\Bigg[
e^{-\rho_h(\pi\rho_h+1)}
\Big\{
3\rho_h^2
\Big[
-\pi e^{\pi\rho_h^2}
\Big(
\rho_h e^{\rho_h}
\Big(
2\,\mathrm{erf}(\sqrt{\pi}\rho_h) \\
&\quad
+ e^{\frac{1}{4\pi}}(1+4\pi)Q^2
\mathrm{erf}\!\left(\frac{2\pi\rho_h+1}{2\sqrt{\pi}}\right)
- 2\rho_h
\Big)
+ Q^2\rho_h^2
- 6Q^2
\Big)
\Big] \\
&\quad
- 2\rho_h e^{\rho_h}
- (4\pi+1)Q^2\rho_h
+ 2Q^2
\Big\} \\
&\quad
+ Q^2 e^{\rho_h(\pi\rho_h+1)} \rho_h^3
\Big[
\pi \rho_h^3 \big(\mathcal{E}_{-2}(\rho_h)-\mathcal{E}_1(\rho_h)\big)
- 3 \mathcal{I}_0(\rho_h)
\Big] + 6Q^2\Bigg]. \label{42}
\end{aligned}
\end{equation}

\begin{figure}[H]
\centering
\includegraphics[width=10cm]{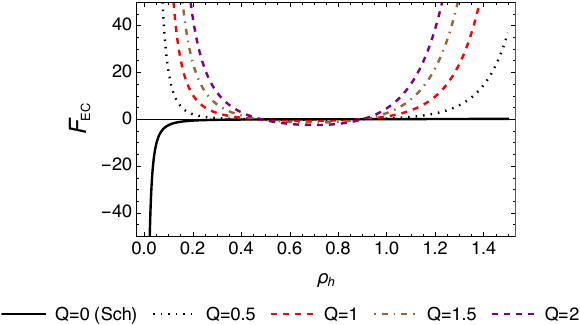}  
\caption{Helmholtz free energy under exponential entropy corrections for different charge $Q$ values.}\label{f16}
\end{figure}

Here, $\text{erf}$ represents the error function. Moreover, due to the presence of a non-integrable contribution in the above expression, this term is isolated and an auxiliary function of the form

\begin{equation}
\mathcal{I}_n(\rho_h)=\int e^{-\pi\rho_h^2} \mathcal{E}_1(\rho_h)\rho_h^{2n}\, d\rho_h, \label{43}
\end{equation}

where $n \in \mathbb{N}^+$. In Fig.~\ref{f16}, under exponential entropy corrections, the exponentially corrected Helmholtz free energy $F$ approaches zero from negative values at $Q=0$ and terminates at zero. This indicates that the system exhibits a unidirectional approach to the reference equilibrium state. For $Q \neq 0$, the energy profile takes on a parabolic structure, encompassing both negative and positive regions, and exhibits an asymmetrical phase transition behavior around zero. As $Q$ increases, this structure becomes tighter, and the transition region narrows. Note that, the contribution of $ \mathcal{I}_0(\rho_h)$, which is not analytically integrable in closed form, has been incorporated into the total Helmholtz free energy via numerical integration methods prior to plotting the results. Also, the same technique will also be applied consistently to similar thermodynamic expressions in subsequent analyses for the contribution of $ \mathcal{I}_1(\rho_h)$ .

By substituting Eqs.\eqref{41} and \eqref{15} into Eq. \eqref{32}, the exponentially corrected internal energy is obtained as

\begin{equation}
\begin{aligned}
\mathcal{U}_{EC}=&\frac{1}{48 \pi \rho_h}
\Bigg[
e^{-\rho_h(\pi\rho_h+1)}
\Big\{
-4\pi e^{\rho_h(\pi\rho_h+1)} \rho_h
\Big[
3\,\mathrm{erf}(\sqrt{\pi}\rho_h) \\
&\quad
+ Q^2\Big(
\rho_h^3\big(\mathcal{E}_{-2}(\rho_h)-\mathcal{E}_1(\rho_h)\big)
- 3 \mathcal{I}_1(\rho_h)
\Big)
\Big] \\
&\quad
- 3(1+2\pi+8\pi^2)Q^2
e^{\frac{(2\pi\rho_h+1)^2}{4\pi}}
\rho_h\,\mathrm{erf}\!\left(\frac{2\pi\rho_h+1}{2\sqrt{\pi}}\right) \\
&\quad
- 6Q^2(\rho_h+4\pi)
+ 12\pi e^{\pi\rho_h^2}
\Big[
\rho_h^2(2e^{\rho_h}+Q^2)
+ 2Q^2
\Big]
\Big\}
\Bigg] \label{44}
\end{aligned}
\end{equation}

\begin{figure}[H]
\centering
\includegraphics[width=10cm]{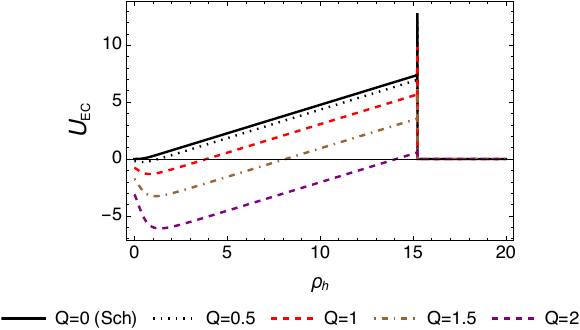}  
\caption{Variation of the exponentially corrected internal energy $\mathcal{U}_{EC}$ with respect to the dimensionless event horizon radius $\rho_{h}$ for different charge values $Q$, under exponential entropy corrections.}\label{f17}
\end{figure}

Fig.\ref{f17} shows the variation of the exponentially corrected internal energy $\mathcal{U}_{EC}$ with the dimensionless event horizon radius ($\rho_{h}$) under exponential entropy corrections for different charge ($Q$) values. Unlike the monotonically increasing behavior in the classical case, the system exhibits a non-monotonic structure, rapidly decreasing after a sharp peak and approaching zero. This indicates a breakdown of stability and a critical transition, with the $Q$ parameter playing a decisive role in the energy transition from negative to positive. 

Substituting exponentially corrected entropy expression \eqref{41} and Eq.\eqref{42} into generic definition of pressure \eqref{22a}, the exponentially corrected pressure is given by

\begin{equation}
\begin{aligned}
\mathcal{P}_{EC}=&\frac{1}{32 \pi^2 \rho_h^6}
e^{-\rho_h}\left(\pi \rho_h^2 + e^{-\pi \rho_h^2}\right) \\
&\quad \times
\Big[
e^{\rho_h}\rho_h^2\left(Q^2 \rho_h^2 \mathcal{E}_1(\rho_h) - 2\right)
+ Q^2\left(\rho_h(-\rho_h^2+\rho_h+6)+6\right)
\Big]. \label{45}
\end{aligned}
\end{equation}

\begin{figure}[H]
\centering
\includegraphics[width=10cm]{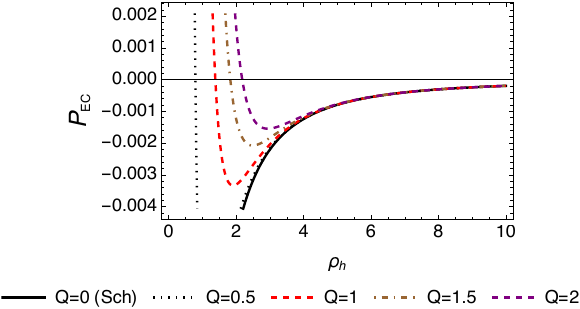}  
\caption{Variation of the exponentially corrected pressure as a function of the dimensionless horizon radius for different charge values.}\label{f18}
\end{figure}

Fig.\ref{f18} shows that the pressure remains negative throughout the parameter range, indicating an effective attractive thermodynamic response. The minimum is highly sensitive to the charge value, reflecting a reshaping of the free energy landscape. Charge introduces a competing repulsive contribution without inducing a sign change, while exponential corrections mainly shift the extremum. The charge effect further diminishes for large dimensionless horizon radius.

In the light of exponentially corrected entropy expression \eqref{41}, the exponentially corrected enthalpy ($\mathcal{H}_{EC}=\mathcal{U}_{EC}+\mathcal{P}_{EC}\mathcal{V}$) becomes 

\begin{equation}
\begin{aligned}
\mathcal{H}_{EC}=&\frac{1}{48\pi}
\Bigg[
-4\pi
\Big(
3\,\mathrm{erf}(\sqrt{\pi}\rho_h)
+ Q^2\rho_h^3\big( \mathcal{E}_{-2}(\rho_h)- \mathcal{E}_1(\rho_h)\big)
- 3Q^2  \mathcal{I}_1(\rho_h)
- 6\rho_h
\Big) \\
&\quad
-3 e^{\frac{1}{4\pi}}(1+2\pi+8\pi^2)Q^2
\mathrm{erf}\!\left(\frac{2\pi\rho_h+1}{2\sqrt{\pi}}\right) \\
&\quad
+\frac{2 e^{-\rho_h}(\pi\rho_h^2+e^{-\pi\rho_h^2})
\Big(
e^{\rho_h}\rho_h^2(Q^2\rho_h^2  \mathcal{E}_1(\rho_h)-2)
+Q^2(\rho_h(-\rho_h^2+\rho_h+6)+6)
\Big)}{\rho_h^3} \\
&\quad
+\frac{6Q^2 e^{-\rho_h(\pi\rho_h+1)}
\Big(-\rho_h+2\pi e^{\pi\rho_h^2}(\rho_h^2+2)-4\pi\Big)}{\rho_h}
\Bigg]. \label{46}
\end{aligned}
\end{equation}

\begin{figure}[H]
\centering
\includegraphics[width=10cm]{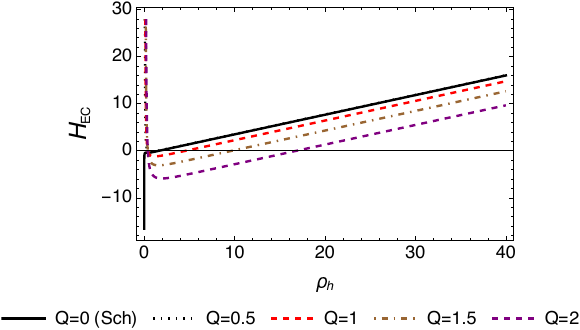}  
\caption{The graph of exponentially corrected enthalpy versus dimensionless horizon parameter with different charge cases.}\label{f19}
\end{figure}

In Fig.\ref{f19} the exponentially corrected enthalpy function shows a monotonic and approximately linear increase with the dimensionless horizon radius, exhibiting a negative vertical shift in the small $\rho_h$ regime at all charge values. This negative negative vertical shift becomes more pronounced as the charge increases, but at the large dimensionless horizon limit, the slope is preserved, and the system transitions to a linear growth regime.

Plugging Eq.\eqref{41}, Eq.\eqref{45} and Eq.\eqref{42} into Eq.\eqref{24g}, the exponentially corrected  Gibbs free energy can be written as

\begin{equation}
\begin{aligned}
\mathcal{G}_{EC}=&\frac{1}{24\pi}
\Bigg[
\frac{e^{-\rho_h(\pi\rho_h+1)}}{\rho_h^3}
\Big\{
\rho_h^2\Big(
-\pi e^{\pi\rho_h^2}
\Big(
e^{\rho_h}\rho_h
\big(
6\,\mathrm{erf}(\sqrt{\pi}\rho_h)
+3e^{\frac{1}{4\pi}}(1+4\pi)Q^2
\mathrm{erf}\!\left(\frac{2\pi\rho_h+1}{2\sqrt{\pi}}\right) \\
&\quad
-6\rho_h
+Q^2\rho_h^3 \mathcal{E}_1(\rho_h)
\big)
- Q^2\big(\rho_h((\rho_h-1)\rho_h+2)+18\big)
\Big)
-3\big(2e^{\rho_h}+(1+4\pi)Q^2\big)
\Big)
+6Q^2
\Big\} \\
&\quad
+ \frac{e^{-\rho_h(\pi\rho_h+1)}}{\rho_h^3}
\left(\pi e^{\pi\rho_h^2}\rho_h^2+1\right)
\Big(
e^{\rho_h}\rho_h^2(Q^2\rho_h^2  \mathcal{E}_1(\rho_h)-2)
+Q^2(\rho_h(-\rho_h^2+\rho_h+6)+6)
\Big) \\
&\quad
-3Q^2  \mathcal{I}_0(\rho_h)
\Bigg]  \label{47}
\end{aligned}
\end{equation}

\begin{figure}[H]
\centering
\includegraphics[width=10cm]{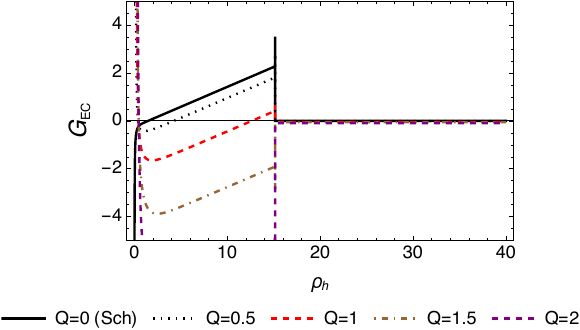}  
\caption{Variation of the exponentially corrected Gibbs free energy $\mathcal{G}_{EC}$ with the dimensionless horizon radius $\rho_h$ for different values of the charge $Q$.}\label{f20}
\end{figure}

Fig.\ref{f20} shows that the exponentially corrected Gibbs free energy $\mathcal{G}_{EC}$ exhibits a non-monotonic behavior similar to that observed for the internal energy. This indicates a modified thermodynamic stability structure, where the charge parameter $Q$ plays a crucial role in the emergence of possible phase transitions.

If we substitute exponentially corrected pressure expression \eqref{45} into Eq.\eqref{25}, the corrected isothermal compressibility reduces to

\begin{equation}
\begin{aligned}
\mathcal{\kappa}_{EC}
= & 48\pi^2 e^{\pi\rho_h^2+\rho_h}\rho_h^6
\Bigg[
\rho_h^3
\Big(
Q^2 e^{\rho_h}(\pi\rho_h^2+1)\mathcal{E}_1(\rho_h)
-2\pi e^{\rho_h}
-\pi Q^2\rho_h \\
&\quad
-2\pi e^{\pi\rho_h^2}(e^{\rho_h}-2Q^2)
+\pi Q^2
\Big)
+ \rho_h^2
Q^2\big(6\pi(2e^{\pi\rho_h^2}+1)-1\big) \\
&\quad
- \rho_h\Big(
4e^{\rho_h}
-12\pi Q^2 e^{\pi\rho_h^2}
-(5+6\pi)Q^2
\Big)
-18Q^2
\Bigg]^{-1} . \label{48}
\end{aligned}
\end{equation}

\begin{figure}[H]
\centering
\includegraphics[width=10cm]{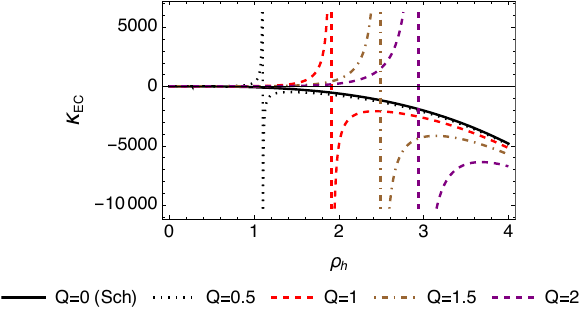}  
\caption{exponential corrected isothermal compressibility behaviors versus the dimensionless black hole horizon for different $Q$ values.}\label{f21}
\end{figure}

Fig.\ref{f21} shows that the isothermal compressibility, under exponential corrections, exhibits singularities for different values of the charge parameter, indicating divergences in the corresponding thermodynamic function.

Finally, in the analysis of the effect of exponential correction on the stability of the black hole, the heat capacity, which is an important quantity, is evaluated using Eq.\eqref{39a}. However, this case it is computed by employing the exponentially corrected internal energy expression given in Eq.\eqref{44}, and the result is obtained as

\begin{equation}
\begin{aligned}
\mathcal{C}_{EC}
=&\frac{1}{4\rho_h^2}
e^{-\rho_h}\left(1-e^{-\pi\rho_h^2}\right)
\Big[
e^{\rho_h}\rho_h^2\left(Q^2\rho_h^2\mathcal{E}_1(\rho_h)+2\right)
- Q^2\left(\rho_h((\rho_h-1)\rho_h+2)+2\right)
\Big] \\
&\quad \times
\Bigg[
\frac{e^{-\rho_h}}{8\pi\rho_h^3}
\Big(
Q^2 e^{\rho_h}\rho_h^4 \mathcal{E}_1(\rho_h)
+4Q^2 e^{\rho_h}\rho_h^3 \mathcal{E}_1(\rho_h)
+2e^{\rho_h}\rho_h^2
+4e^{\rho_h}\rho_h \\
&\qquad
- Q^2\rho_h^3
- Q^2(3\rho_h^2-2\rho_h+2)
\Big)
-\frac{e^{-\rho_h}}{8\pi\rho_h^3}
\Big(
Q^2 e^{\rho_h}\rho_h^4 \mathcal{E}_1(\rho_h)
+2e^{\rho_h}\rho_h^2 \\
&\qquad
- Q^2(\rho_h^3-\rho_h^2+2\rho_h+2)
\Big)
-\frac{3e^{-\rho_h}}{8\pi\rho_h^4}
\Big(
Q^2 e^{\rho_h}\rho_h^4 \mathcal{E}_1(\rho_h)
+2e^{\rho_h}\rho_h^2 \\
&\qquad
- Q^2(\rho_h^3-\rho_h^2+2\rho_h+2)
\Big)
\Bigg]^{-1}  \label{49}
\end{aligned}
\end{equation}

\begin{figure}[H]
\centering
\includegraphics[width=10cm]{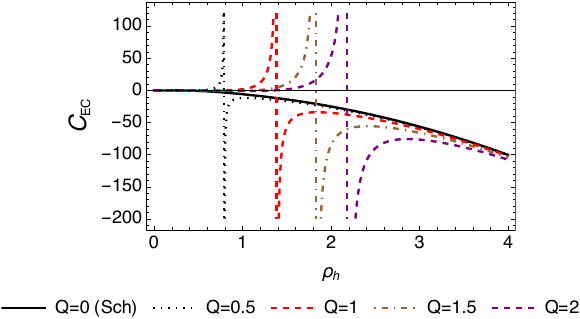}  
\caption{Heat capacity as a function of the dimensionless horizon radius $\rho_h$ for different values of the charge parameter $Q$ under exponential corrections.}\label{f22}
\end{figure}

The graphs presented in Fig.\ref{f22} illustrate the $\rho_h-$ dependent behavior of the $\mathcal{C}_{EC}$ under different $Q$ parameters. When exponential corrections are added to the black hole entropy, the thermodynamic stability structure undergoes a transformation. As shown in Fig.\ref{f22}, the resulting stability region now diverges at specific $\rho_h$  points, as in the uncorrected and logarithmic cases, and these divergence points are strongly dependent on the $Q$ value, as in the uncorrected or logarithmic correction cases.

\section{Results and Discussions}

In this work, we have investigated the thermodynamic behaviour of a black hole solution of general relativity sourced by nonlinear Yukawa-type electrodynamics in the presence of quantum-corrected entropy contributions. In particular,  the analysis  is based on the inclusion of both logarithmic and exponential corrections to the standard Bekenstein–Hawking entropy, which are generally expected to arise from quantum gravitational effects and thermal fluctuations near the black hole horizon. These corrections provide a more comprehensive description of the microscopic structure of black hole thermodynamics beyond the semiclassical approximation.

Starting from the modified entropy expressions, the corresponding thermodynamic quantities were derived and the effects of these corrections on the physical properties and stability structure of the Yukawa black hole spacetime were examined. Special attention has been devoted to the behavior of key thermodynamic parameters, including the temperature, heat capacity, Helmholtz free energy, and related stability criteria. The effect of the charge  have been analyzed in detail in order to determine their role in the thermodynamic evolution of the system. Furthermore, the thermodynamic characteristics of the model have been explored extensively through graphical analysis. The corresponding plots were generated primarily with respect to the charge parameter of the Yukawa black hole, allowing us to investigate the influence of charge on the corrected thermodynamic quantities and possible phase transition behavior. Our results indicate that the inclusion of logarithmic and exponential entropy corrections modifies the thermodynamic response of the system, especially in regions associated with critical behavior and stability transitions.

The physical information obtained from the corrected thermodynamic analysis lies in the different manner in which the two entropy corrections affect the Yukawa-screened system. The logarithmic correction generally preserves the qualitative form of the classical thermodynamic potentials and produces its most visible effects through moderate displacements of the critical points. The exponential correction, by contrast, generates more pronounced changes in the small-horizon regime and leads to qualitatively different charge dependences in some of the energy functions. Therefore, the two correction schemes cannot be regarded as equivalent modifications of the Bekenstein–Hawking entropy in this background: the logarithmic term mainly perturbs the existing classical structure, whereas the exponential term can substantially reorganize the thermodynamic response of the system.

The comparison of the classical, logarithmically corrected, and exponentially corrected cases reveals a common large-horizon tendency together with important differences in the small-horizon regime. For sufficiently large values of the dimensionless horizon radius, the charge-dependent curves approach the Schwarzschild behaviour, indicating that both the nonlinear electromagnetic contribution and the entropy corrections become progressively less influential. At small horizon radii, however, the corrected quantities remain sensitive to the charge and to the form of the entropy correction. The logarithmic correction generally produces controlled shifts relative to the classical curves, whereas the exponential correction leads to a more pronounced restructuring of the internal energy, enthalpy, and Gibbs free energy. In particular, the charge dependence of the enthalpy is reversed, while the Schwarzschild configuration yields the largest values of the internal and Gibbs energies over part of the relevant small-horizon domain. These differences show that the choice of entropy correction determines not only the magnitude of the thermal quantities but, in certain cases, also their qualitative response to the nonlinear electromagnetic charge.

The stability indicators provide the clearest distinction among the three thermodynamic descriptions. For the isothermal compressibility, the logarithmic correction shifts the divergence points toward smaller horizon radii for all charged configurations, thereby changing the location at which the system becomes highly sensitive to pressure variations. The exponential correction, on the other hand, largely preserves the qualitative structure of the classical compressibility, although the positions of its singularities remain charge dependent. The heat capacity displays a different response: both entropy corrections shift its divergence points toward smaller horizon radii relative to the uncorrected case. These results show that the corrections affect different stability indicators in different ways. The logarithmic term has a more visible influence on the compressibility boundary, whereas both correction schemes modify the heat-capacity boundary separating locally stable and unstable branches.

Overall, the logarithmic and exponential entropy corrections are considered to probe two physically distinct departures from the classical Bekenstein--Hawking description. The logarithmic term represents the leading effect of thermal fluctuations around equilibrium and therefore tests whether the thermodynamic and stability properties induced by Yukawa screening remain robust against such fluctuations. The exponential correction, in contrast, probes non-perturbative entropy effects that are strongly suppressed in the large-entropy regime but may become significant for small black holes. Studying both corrections therefore allows us to examine the robustness of Yukawa-screened black hole thermodynamics beyond the classical description, particularly in the small-horizon regime where these effects are expected to be most relevant.

The electrogeodesic formulation presented here also provides a natural starting point for a more comprehensive dynamical investigation. In future work, a systematic analysis of charged-particle orbital dynamics, including a comparison with the Reissner--Nordstr\"om spacetime, will be complemented by a perturbative stability analysis through the computation of quasinormal modes.

\begin{acknowledgments}
The author expresses his sincere gratitude to Prof. Dr. Mustafa Halilsoy, anonymous reviewers and Dr. Huriye Gursel for their insightful comments and constructive suggestions, which have significantly improved the quality and clarity of this manuscript.
\end{acknowledgments}

\bibliographystyle{unsrt}

\end{document}